\documentclass[12pt]{article}

\pdfoutput=1

\usepackage{graphics}
\usepackage{amssymb}
\usepackage{amsmath}

\newcommand{\rf}[1]{(\ref{#1})}
\newcommand{\beq}{\begin{equation}}
\newcommand{\eeq}{\end{equation}}
\newcommand{\bea}{\begin{eqnarray}}
\newcommand{\eea}{\end{eqnarray}}

\newcommand{\e}{\mbox{e}}

\newcommand{\lam}{\lambda}
\newcommand{\Lam}{\Lambda}
\newcommand{\bt}{\beta}
\newcommand{\al}{\alpha}

\newcommand{\m}{\mu}

\newcommand{\kp}{\kappa}
\newcommand{\vph}{\varphi}

\newcommand{\oh}{\frac{1}{2}}

\newcommand{\tr}{\mathrm{tr}\,}
\newcommand{\ra}{\rangle}
\newcommand{\la}{\langle}

\newcommand{\cT}{{\cal T}}

\newcommand{\cZ}{{\cal Z}}

\newcommand{\kpdt}{{\kp_{{\rm dt}}}}

\newcommand{\rmd}{\mathrm{d}}

\begin{document}

\begin{center}
\vspace{24pt}
{ \large \bf Scale-dependent Hausdorff dimensions in 2d gravity}

\vspace{30pt}

{\sl J. Ambj\o rn}$\,^{a,c}$,
{\sl T. Budd}$\,^{a}$,
and {\sl Y. Watabiki}$\,^{b}$

\vspace{48pt}
{\footnotesize

$^a$~The Niels Bohr Institute, Copenhagen University\\
Blegdamsvej 17, DK-2100 Copenhagen \O , Denmark.\\
{ email: ambjorn@nbi.dk, budd@nbi.dk}\\

\vspace{10pt}

$^b$~Tokyo Institute of Technology,\\ 
Dept. of Physics, High Energy Theory Group,\\ 
2-12-1 Oh-okayama, Meguro-ku, Tokyo 152-8551, Japan\\
{email: watabiki@th.phys.titech.ac.jp}

\vspace{10pt}

$^c$~Institute for Mathematics, Astrophysics and Particle Physics (IMAPP)\\ 
Radbaud University Nijmegen,\\ 
Heyendaalseweg 135, 6525 AJ, Nijmegen, The Netherlands 

}
\vspace{96pt}
\end{center}

%\addtolength{\baselineskip}{0.20\baselineskip}
%\vspace{2cm}

\begin{center}
{\bf Abstract}
\end{center}

By appropriate scaling of coupling constants a one-parameter family of ensembles of two-dimensional geometries is obtained, which interpolates between the ensembles of (generalized) causal dynamical triangulations and ordinary dynamical triangulations.
We study the fractal properties of the associated continuum geometries and identify both global and local Hausdorff dimensions.

\vspace{12pt}
\noindent

\vspace{24pt}

\newpage

\section{Introduction}\label{intro}

In dynamical triangulations (DT) the path integral of two-dimensional Euclidean quantum gravity
is discretized by summing over equilateral triangulations.
If the topology of the two-dimensional manifold is kept fixed, 
the Einstein curvature term in the action is trivial 
(being a topological invariant) and can be safely ignored. 
In that case the DT path integral takes the form
\beq\label{1.1}
Z = \sum_{T \in \cT} \frac{1}{C_T} \;\e^{-\m N_T},
\eeq
where the sum is over all combinatorial triangulations of the desired topology, $C_T$ is the order of its automorphism group, $N_T$ its number of triangles, and $\mu$ is a coupling constant. The matrix 
model representation of eq.\  \rf{1.1} is 
\beq\label{1.2}
\cZ = \int d \phi \; e^{-N \tr \left(\oh \phi^2 -\frac{\kpdt}{3} \phi^3\right)},
~~~Z= \log \cZ,~~~~\kpdt = e^{-\m},
\eeq
where the integration is over the Hermitian $N\times N$ matrices $\phi$. 
The partition function $Z$ allows for an expansion in $1/N^2$ and the Feynman diagrams contributing to the coefficient of $N^{\chi}$ are precisely the cubic graphs dual to triangulations of Euler characteristic $\chi$ appearing in \rf{1.1}.
In this paper we will only deal with the leading term 
in the $1/N$ expansion, i.e.\ 
cubic graphs with spherical topology (or, in cases where 
a boundary is present, the topology of the disk). 

The lattice action has a critical point at $\kpdt\to(\kpdt)_c$ and the corresponding continuum limit can be identified with quantum Liouville
field theory with central charge $c_{ \rm liouville}=26$. 
Universality of the scaling 
limit ensures that one obtains the same continuum limit 
for any potential
\beq\label{1.3}
V(\phi) = \frac{1}{g}\Big(\oh \phi^2 -\sum_{n} \kp_n \phi^n\Big),
\eeq
instead of the cubic potential used in \rf{1.2}, provided\footnote{Some
 mild constraint has to be imposed if an infinite number of $\kp_n \neq 0$.}  
the $\kp_n \geq 0$ and at least one $\kp_n >0$ for $n \geq 3$. Thus 
one can replace the set of triangulations $\cT$ in \rf{1.1} (or rather the dual cubic graphs) with
a much larger class of graphs if desired \cite{ajm}. 
Independent of the precise class of graphs and as long as one keeps the couplings in the potential fixed as $N\to\infty$, the geometry of a randomly sampled very large graph has a number of universal properties, one of these
being that its fractal dimension is $d_h=4$
rather than the naively expected $d_h=2$ \cite{kawai,watabiki,aw,ajw}.
In the following we will refer to this universal continuum limit, which in the mathematical literature is known as the Brownian map \cite{brownianmap}, as the DT continuum limit.  

However, it is possible to define a different scaling limit
for which $d_h=2$ by scaling the coupling $g$ in 
\rf{1.3} non-trivially as function of $N$. Scaling  
$g \to 0$ as $g = G a^3$ while keeping the continuum volume $\propto N a^2$ fixed, where $a$ may be interpreted as the length of a link in the graph,
leads to a scaling limit known as generalized causal dynamical triangulations (GCDT) 
\cite{gcdt}. 
It generalizes the original model of CDT in the continuum \cite{al}, which arises as the $G\to 0$ limit of GCDT and is different from Liouville quantum gravity.\footnote{Instead, continuum CDT was shown to correspond to two-dimensional Ho\v{r}ava-Lifshitz gravity \cite{aggsw}.} 

In \cite{ab} the GCDT was shown to arise explicitly as a scaling limit of random quadrangulations with a fixed number of local maxima of the distance function to a distinguished vertex.
These quadrangulations were shown to be in bijection with general planar graphs with a fixed number of faces has been given, of which the continuum limit is therefore also described by GCDT.

The purpose of the present article is to investigate scalings that interpolate between the  
two ``extremes'' mentioned above, DT and GCDT. For that purpose we will restrict our
attention to a simple potential \rf{1.3} of the form 
\beq\label{1.4}
V_0(\phi) = 
\frac{1}{g} \Big( -\kp \phi + \oh \phi^2 -\frac{\kp}{3} \phi^3\Big).
\eeq

\section{The scaling limit}

The disk amplitude for a general potential \rf{1.3} has the form:
\beq\label{2.1}
w(z) = \frac{1}{2} \Big(V'(z)- A(z)\sqrt{(z-c)(z-d)}\Big),
\eeq
where $V(z)$ is the potential \rf{1.3} with the matrix $\phi$ replaced
by the complex number $z$, and $V'(z)$ denotes the derivative with 
respect to $z$. The polynomial $A(z)$ and the numbers $b,c,d$ (with $c\geq d$) are uniquely determined by the requirement that 
$w(z) \to 1/z$ for $|z| \to \infty$. For the potential \rf{1.4} we write
\beq\label{2.2}
w(z) = \frac{1}{2g} \Big(-\kp +z-\kp z^2 + \kp (z-b) \sqrt{(z-c)(z-d)}\Big),
\eeq   
where $z = e^{\lam_{B}}$, $\lam_B$ having  the interpretation as  
a boundary cosmological constant associated with the disk-boundary 
for positive $z$.

For a fixed $g$ the critical point $\kp_c$ is determined by the
condition that $b(\kp_c) = c(\kp_c)$, where $b(\kp),c(\kp)$ and $d(\kp)$
are determined by the required asymptotics of $w(z)$, as mentioned.
The solution can be written as follows (denoting $b(\kp_c)$ by $b_c$ etc.):
\beq\label{2.3}
b_c = c_c,~~~(b_c-d_c)^3 = \frac{32g}{\kp_c},~~~~~
\kp_c b_c = \oh+\oh (4\kp_c^2\,g)^{1/3}
\eeq
\beq\label{2.4}
\left(1-4\kp_c^2\right)^{3/2}=3^{3/2}\,4\kp_c^2\, g.
\eeq     
From these equations one observes that if we scale 
$g$ to 0  as $g= G\, a^3$ one obtains to lowest order 
\beq\label{2.5}
\kp^2_c = \frac{1}{4} -\frac{3}{4} G^{2/3} a^2,~~~
b_c = \oh+ \frac{3}{4} G^{1/3}a,~~~b_c-d_c= 4 G^{1/3} a,
\eeq
as discussed in \cite{gcdt}.

We will now show how to understand this scaling limit in 
a simple way which also allows us to define more 
general scaling limits.

\begin{figure}
%\vspace{-0.5cm}
\centerline{\scalebox{0.35}{\rotatebox{0}{\includegraphics{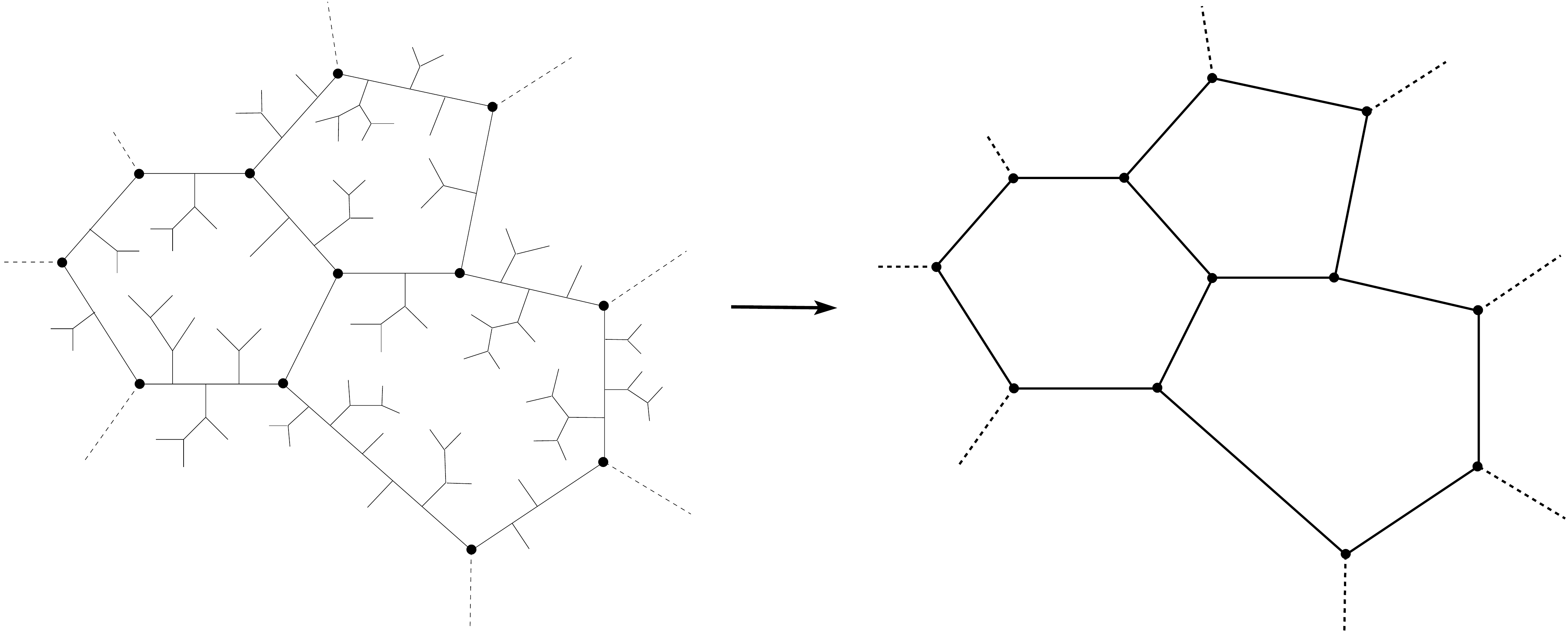}}}}
\caption{Left: A typical graph generated by the action \rf{1.4}. Right:
Summing over the trees one obtains a cubic ``skeleton'' graph.}
\label{tadpoles}
\end{figure}

\begin{figure}
%\vspace{-0.5cm}
\centerline{\scalebox{0.5}{\rotatebox{0}{\includegraphics{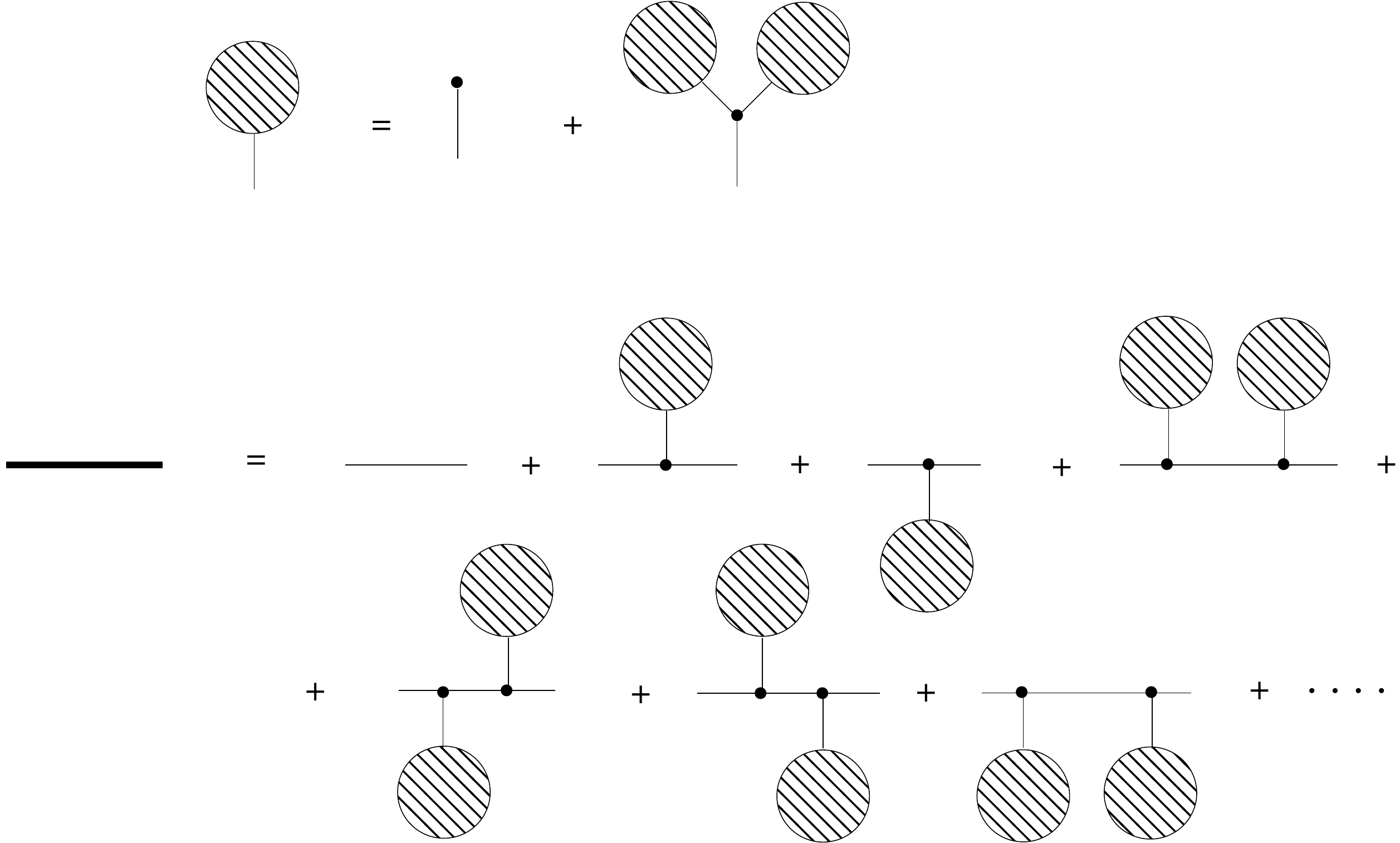}}}}
\caption{Top figure: the graphic equation defining the summation 
over all rooted trees. Bottom figure: The graphic equation for 
the dressed propagator which appears in Fig.\ \ref{tadpoles} after summing 
over all tree-outgrows.}
\label{propagator}
\end{figure}

Consider the partition function \rf{1.2} with the potential 
$V_0(\phi)$ defined by eq.\ \rf{1.4}. Expanding the exponential 
in powers of $\kp$ and performing the Gaussian integrals 
can be viewed as generating a certain set of graphs. We can view 
these graphs as $\phi^3$ graphs decorated with tadpoles coming
from the linear $\phi$ term, see Fig. \ref{tadpoles}. 
The shift in integration variables by 
\beq\label{2.6}
\phi = \vph+ \al(\kp),~~~~~~\al(\kp) = \frac{1-\sqrt{1-4\kp^2}}{2\kp}
\eeq
eliminates the tadpole term:
\beq\label{2.7}
V_1(\vph) =V_0(\vph +\al(\kp)) = \frac{1}{g} 
\left( \frac{\sqrt{1-4\kp^2}}{2}\,\vph^2-
\frac{\kp}{3} \,\vph^3 \right) +{\rm const.}
\eeq 
The constant will play no role when we calculate expectation
values of observables. In terms of 
graphs it means that we are introducing a ``dressed'' 
propagator  by first summing over all tadpole terms.
This is illustrated in Fig.\ \ref{tadpoles} and 
in more detail in Fig.\ \ref{propagator}. We call the graph left
after summing over the tadpole terms for the skeleton 
graph. 

The constant $\al(\kp)$
is precisely the summation over all connected planar, rooted tree diagrams,
where each line has weight $g$ and each vertex weight $\kp/g$, as dictated
by the action $V_0(\phi)$. This summation is shown in the upper part 
of Fig.\ \ref{propagator}. Next, the lines remaining 
in the skeleton graph have  the weight $g\bt(\kp)$ as 
illustrated in the lower part of Fig.\ \ref{propagator}, where
\beq\label{2.7a}
\bt(\kp) = 1 +  [2\kp \al(\kp)] +  [2\kp \al(\kp)]^2 + \cdots 
= \frac{1}{\sqrt{1-4\kp^2}}.
\eeq
This explains the form of $V_1(\vph)$ from the point of view 
of graph re-summation. Both for a graph $G$ generated from   
$V_0(\phi)$ or for its corresponding skeleton graph  generated 
from $V_1(\vph)$, the power of $g$ associated with the graph is 
\beq\label{2.7b}
g^{F(G)-2},~~~F(G)=\mbox{number  of faces of $G$}.
\eeq    
Thus it is clear that when we take $g \to 0$ we will suppress
the number of faces of the graph.

Finally we can perform a rescaling 
\beq\label{2.8}
\vph = \sqrt{\frac{g}{\sqrt{1-4\kp^2}}} \; \Phi,
\eeq
such that 
\beq\label{2.9}
V_1(\vph) = V_2(\Phi) = \oh\, \Phi^2 -\frac{\kpdt}{3}\,\Phi^3,~~~~~~
\kpdt= \frac{\sqrt{g}\, \kp}{(1-4\kp^2)^{3/4}}.   
\eeq
In a graph $G$ this change of variables corresponds to absorbing 
the weight $g/\sqrt{1-4\kp^2}$
given to each link by $V_1(\vph)$ into  the two vertices associated with 
the link. Thus the weight of each link is 1, but the coupling 
constant associated with a vertex is changed from $\kp/g$ to $\kpdt$. 
Again, this rescaling will not affect expectation values 
of observables. The potential $V_2(\Phi)$ is the standard
potential used to represent dynamical triangulations  
using matrix models, and the critical coupling is known
to be $(\kpdt)_c = 3^{1/4}/6$. Using eq.\ \rf{2.4} we can write:
\beq\label{2.10}
\frac{\kpdt}{(\kpdt)_c} = \frac{\kp}{\kp_c}\, 
\left( \frac{1-4\kp_c^2}{1-4\kp^2}  \right)^{3/4}.
\eeq

This formula captures the  different ways one can take the scaling 
limit for the model given by  $V_0(\phi)$.
The  tree sub-graphs shown in the left part of 
Fig.\ \ref{tadpoles} and in the top part of Fig.\ \ref{propagator}
have the partition function $\al(\kp)$ given in \rf{2.6}, and become critical 
for $\kp^2 \to 1/4$. The average number of vertices in a tree is 
\beq\label{2.11}
\la n \ra_\kp = \frac{\kp}{\al} \,\frac{d \al}{d \kp} 
\sim \frac{1}{\sqrt{1-4\kp^2}}.
\eeq
Similarly, the average number of vertices $\la n_{prop}\ra_\kp$ associated 
with trees  attached to the dressed ``propagator''
shown in Fig.\ \ref{propagator} diverges when  $\kp^2 \to 1/4$.  
The partition function for the number of such vertices is $\bt(\kp)$
defined in \rf{2.7a}, and  

\beq\label{2.12}
\la n_{prop} \ra_{\kp} = \frac{\kp}{\bt}\, \frac{d\bt}{d \kp}  
\sim \frac{1}{1-4\kp^2}.
\eeq
Thus the total average number of trees attached to a dressed propagator 
is proportional to $1/\sqrt{1-4\kp^2}$.
Since $\kp \leq  \kp_c$ we conclude from eq.\ \rf{2.4} that 
if we keep $g >0$ fixed when taking the scaling limit
$\kp \to \kp_c$ the trees will not be critical since 
$\kp_c^2 < 1/4$. Thus the trees can  basically be ignored 
in the scaling limit $\kp\to \kp_c$.  Eq.\ \rf{2.10} captures this: for 
$g>0$ it tells us that 
\beq\label{2.13}
\frac{\kpdt}{(\kpdt)_c} = \frac{\kp}{\kp_c}\, \Big( 1+ O( (\kp_c-\kp)^2)\Big).
\eeq
The critical behavior is thus the standard one of DT and the graphs 
responsible for this are the standard $\phi^3$ graphs. In this scaling 
limit we write
\beq\label{2.14} 
\frac{\kp}{\kp_c} = 1-a^2 \Lam + o(a^2),
\eeq
where $\Lam$ may be interpreted as the cosmological 
constant. 

Clearly, to obtain  a different scaling limit we have to scale 
$g$ to zero when $a \to 0$. The GCDT limit was obtained 
by the scaling $g = G \,a^3$, and using \rf{2.5} we obtain from \rf{2.10}
\beq\label{2.15}
\frac{\kpdt}{(\kpdt)_c} = \frac{\kp}{\kp_c}\,
\left(\frac{ 3G^{2/3}}{ 3G^{2/3} +2\Lam}\right)^{3/4}.
\eeq
This shows that $\kpdt$ does not become critical as $\kp\to\kp_c$. Thus there is 
only a finite average number of vertices and links and faces  
in the skeleton graph. The 
critical behavior for $\kp \to \kp_c$ is entirely determined
by the trees. 

In order to obtain a new limit, let us consider the scaling
\beq\label{2.16}
g = G_\al a^\al,~~~~0<\al < 3,
\eeq
while maintaining \rf{2.14}, which states that 
we view the total number of vertices as proportional with 
the continuum area of the graph. With this scaling 
of $g$ we obtain from \rf{2.5} and \rf{2.10}
\beq\label{2.17}
\frac{\kpdt}{(\kpdt)_c} = 
\left( 1 -\frac{\Lam}{2 G_\al^{2/3}}\, a^{2-2\al/3} + 
o( a^{2-2\al/3})\right).
\eeq
By a scaling like \rf{2.16} we thus obtain that both the trees and 
the skeleton graphs are critical. The average number of graph vertices
per link in the skeleton graph is  
\beq\label{2.18}
\la n_{prop} \ra_\kp \sim \frac{1}{1-4\kp^2} \sim \frac{1}{a^{2\al/3}},
\eeq
while the average number of skeleton vertices\footnote{For a planar 
$\phi^3$ graph we have $3V=2L$ and $3F-L=6$, where $V,L$ and $F$ denotes 
the number of vertices, links and faces in the graph.}  will be 
\beq\label{2.19}
\la n_{skel} \ra_{\kp} \sim \frac{1}{a^{2-2\al/3}},
\eeq
implying that the  total number vertices in a typical graph 
scales as $a^{-2}$, in accordance with \rf{2.14}.
Notice also that the average length of a skeleton link scales as $1/\sqrt{1-4\kappa^2}\sim a^{-\alpha/3}$.

In order to study the fractal properties of this ensemble of 
graphs one may calculated the so-called two-point function \cite{aw,ajw},
i.e.\ the partition function with two distinguished vertices 
separated by a given link distance $r$. It can be calculated using 
the methods in \cite{ab} or \cite{aw,ajw}. For small $g$, $\kp$ close to $\kp_c$
and $r\gg g^{-1/3}$ one finds (up to numerical constants)
\beq\label{2.19a}
Z_\mu(r,g) \sim \Big(g^{2/3} \mu\Big)^{3/4} \; 
\frac{\cosh \Big((g^{2/3} \mu)^{1/4} \, r\Big)}{ 
\sinh^3 \Big((g^{2/3} \mu)^{1/4} \, r\Big)},~~~~\mu = \frac{\kp_c-\kp}{\kp_c}.
\eeq 
Let us now take the scaling limit prescribed by eqs.\ 
\rf{2.14}  and \rf{2.16}. Insisting on keeping $G_\al$ and $\Lam$ fixed
and eliminating the scaling parameter $a$ in favor of $\mu$ leads to
\beq\label{2.19b}
Z_\mu(r) \sim (K \mu^{1/d_H})^3\, 
\frac{\cosh (K \mu^{1/d_H} r)}{ \sinh^3 (K \mu^{1/d_H} r)},~~~
K  = \Big(\frac{G}{\Lam^{\al/2}}\Big)^{1/6},
\eeq
which holds for $r \gg \mu^{-\alpha/6}$ and where\footnote{A similar scaling was anticipated in \cite{threepoint}, Section 6.}
\beq\label{2.20}
d_H = \frac{4}{1+\al/3}.
\eeq
Since $(1-\mu)$ is a generating variable for the number $N$ of vertices in the graph, we find that the (canonical) two-point function $Z_N(r)$ for fixed $N$ is of the form ,
\beq
Z_N(r) \sim F(r/N^{1/d_H})\quad \text{for }r \gg N^{\alpha/6}
\eeq 
where $F(R)$ is some function that goes to zero fast for $R\gg 1$.
In particular, this implies that the average distance $\langle r\rangle$ between arbitrary vertices is of the order $N^{1/d_H}$. 
For this reason one may call the exponent $d_H$ the ``global'' Hausdorff dimension \cite{aw,ajw}. 

\begin{figure}
\centerline{\scalebox{1.4}{\rotatebox{0}{\includegraphics{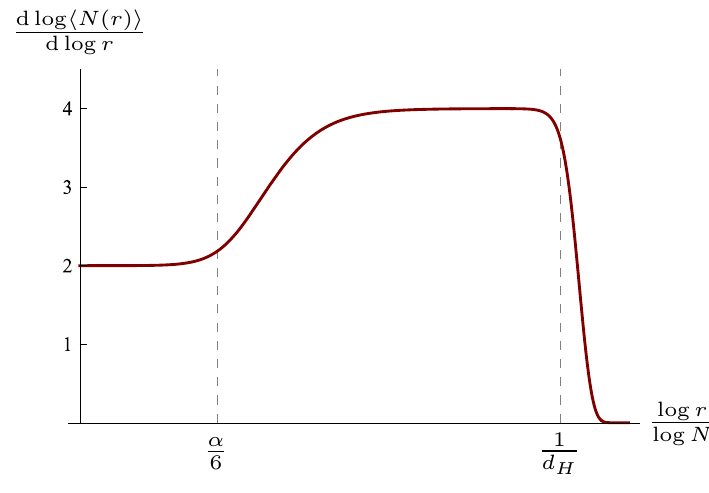}}}}
\caption{The scale-dependent Hausdorff dimension for $0<\alpha<3$.}
\label{dimension}
\end{figure}

The dimension $d_H$ should be contrasted with the ``local'' Hausdorff dimension $d_h$, which is associated with the growth $V(R)\sim R^{d_h}$ for small $R$ of the expected volume $V(R)$ of a disk as function of its radius $R$ within a fixed continuous geometry.
In order to associate a well-defined Hausdorff dimension $d_h$ to our ensemble of graphs, one first has to specify how to scale the distance $r$ in the continuum limit.
If one defines the continuum distance $R$ as $R = r a^{2/d_H}$ eq. \rf{2.19b} reduces to
\beq
Z_\mu(r) \sim \frac{\cosh(\Lambda^{1/4}G_{\alpha}^{1/6} R)}{\sinh^3(\Lambda^{1/4}G_{\alpha}^{1/6} R)},
\eeq
which, up to the factor $G_{\alpha}^{1/6}$ is precisely the DT two-point function and therefore $d_h=4$.
If, on the other hand, we scale $R = r a^{\alpha/3}$, the skeleton edges will maintain a finite length in the continuum, meaning that the local Hausdorff dimension is determined by that of the trees, i.e. $d_h=2$.

These observations can be summarized by looking at the ``scale-dependent'' Hausdorff dimension
\beq
d_h(N;r) := \frac{\rmd \log\langle N(r)\rangle}{\rmd \log(r)}
\eeq
for a fixed large $N$, where $\langle N(r)\rangle$ is the expected number of vertices within graph distance $r$ from a randomly chosen vertex. 
A qualitative plot of $d_h(N;r)$ as function of $\log(r)/\log(N)$ for some $0<\alpha<3$ is shown in Fig.\ \ref{dimension}.
The local Hausdorff dimensions, $d_h=2$ and $d_h=4$, appear as plateaus, while the global Hausdorff dimension corresponds to the scale at which $d_h(r)$ drops to zero.

\section{Discussion} 

The partition function 
\beq\label{3.1}
\cZ = \int d \phi \; 
\e^{- \frac{N}{g} \tr [-\kp \phi + \oh \phi^2 -\frac{\kp}{3} \phi^3]}
\eeq
generates a statistical ensemble of graphs of the kind shown in 
Fig.\ \ref{tadpoles}. 
The coupling constant $g$ in the action \rf{1.4} can be viewed as
the   temperature $k T$ of this statistical system. Thus a scaling limit 
where $g > 0$ corresponds to a finite temperature, and this finite temperature 
limit can be identified with the standard scaling limit of 2d Euclidean 
quantum gravity: i.e. the typical geometry of the ensemble is fractal
with Hausdorff dimension $d_h=4$.  

The scaling limit $g \to 0$ has certain analogues with the  
annealing, quenching and tempering of alloys and metals, in the sense that the 
precise way we take this zero temperature limit decides 
the smoothness of a typical geometry dominating at zero 
temperature. The parameter controlling this is the 
exponent $\al$ when writing $g= G_\al a^{\al}$, $0 \leq \al \leq 3$,
which tells us how ``fast'' we cool to zero temperature. 
The so-called global Hausdorff dimension $d_H$ of a typical graph 
in such an $\al$-ensemble is given by
\beq
d_H = \frac{4}{1+\alpha/3},
\eeq
while, depending on the chosen scaling of the geodesic distance, the local Hausdorff dimension 
is either $d_h=4$ or $d_h=2$.

Here we have only considered the simplest situation, that 
of spherical (or disk) topology of the graphs and the associated
geometries. It remains to be seen if there
exists a complete perturbative expansion in topology and 
in number of boundaries for an arbitrary value of $\al$, as 
is the case in the two limits $\al =0$ and $\al = 3$.

\subsection*{Acknowledgments}

The authors acknowledge support from the ERC-Advance grant 291092,
``Exploring the Quantum Universe'' (EQU). JA acknowledges support
of FNU, the Free Danish Research Council, from the grant
``quantum gravity and the role of black holes''.   
In addition JA was supported 
in part by Perimeter Institute of Theoretical Physics.
Research at Perimeter Institute is supported by the Government of Canada
through Industry Canada and by the Province of Ontario through the 
Ministry of Economic Development \& Innovation.

\end{document}